\documentclass[twocolumn]{aastex63}
\usepackage{amsmath}
\usepackage{svg}

\DeclareMathOperator*{\argmin}{arg\,min}

\submitjournal{ApJ}
\received{12 Oct 2023}
\revised{22 Dec 2023}
\accepted{03 Jan 2024}

\shorttitle{Probing Type Ia Supernova Explosions}
\graphicspath{{./}{figures/}}

\begin{document}

\title{Probing the Diversity of Type Ia Supernova Light Curves in the Open Supernova Catalog}

\author[0009-0003-2660-2018]{Chang Bi}
\affiliation{Department of Mathematics and Statistics, University of Victoria, 3800 Finnerty Road, V8P 5C2, Victoria, BC, Canada}
\affiliation{National Research Council of Canada, Herzberg Astronomy \& Astrophysics
Research Centre,\\
5071 West Saanich Road, Victoria, BC V9E 2E7, Canada}

\author[0000-0003-1428-5775]{Tyrone E.~Woods}
\affiliation{National Research Council of Canada, Herzberg Astronomy \& Astrophysics
Research Centre,\\
5071 West Saanich Road, Victoria, BC V9E 2E7, Canada}
\affiliation{Department of Physics and Astronomy, Allen Building, 30A Sifton Rd, University of Manitoba, Winnipeg MB  R3T 2N2, Canada}

\author[0000-0003-2239-7988]{S{\'e}bastien Fabbro}
\affiliation{National Research Council of Canada, Herzberg Astronomy \& Astrophysics
Research Centre,\\
5071 West Saanich Road, Victoria, BC V9E 2E7, Canada}

\begin{abstract}

The ever-growing sample of observed supernovae enhances our capacity for comprehensive supernova population studies, providing a richer dataset for understanding the diverse characteristics of Type Ia supernovae and possibly that of their progenitors. Here, we present a data-driven analysis of observed  Type Ia supernova photometric light curves collected in the Open Supernova Catalog. Where available, we add the environmental information from the host galaxy. We focus on identifying sub-classes of Type Ia supernovae without imposing the pre-defined sub-classes found in the literature to date. To do so, we employ an implicit-rank minimizing autoencoder neural network for developing low-dimensional data representations, providing a compact representation of the supernova light curve diversity. When we analyze light curves alone, we find that one of our resulting latent variables is strongly correlated with redshift, allowing us to approximately ``de-redshift'' the other latent variables describing each event. After doing so, we find that three of our latent variables account for $\sim$95\% of the variance in our sample, and provide a natural separation between 91T and 91bg thermonuclear supernovae. Of note, the 02cx subclass is not unambiguously delineated from the 91bg sample in our results, nor do either the over-luminous 91T or the under-luminous 91bg/02cx samples form a clearly distinct population from the broader sample of ``other'' SN Ia events. We identify the physical characteristics of supernova light curves which best distinguish SNe 91T from SNe 91bg \& 02cx, and discuss prospects for future refinements and applications to other classes of supernovae as well as other transients.

\end{abstract}


\section{Introduction}
\label{sec:intro}

Supernovae are easily among the most optically-luminous astrophysical transients \citep{LSSTScienceBook}. These explosive endpoints of stellar evolution shape their surrounding interstellar gas, enriching it as a primary source of heavy elements in the Universe \citep{Johnson2019}, and carving out the ``hot'' phase of the interstellar medium \citep{Chevalier1977}. Their expanding shock waves are also powerful cosmic ray accelerators \citep{BZ34,BE87}, and may trigger the formation of new stars \citep[e.g.,][]{Kothes01}.

\begin{table*}[]
\centering
\caption{Stastistics for the Open Supernova Catalogue light curve data considered for this analysis}
\label{tab:datacuts}
\begin{tabular}{|l|r|r|}
\hline
Condition          & Measurements & Supernovae \\ \hline
initial sample     & 1,273,312    & 22,528     \\ \hline
contains photometry data & 1,273,312    & 14,468     \\ \hline
pass quality cuts  & 1,273,185    & 14,460     \\ \hline
has host information &   954,824    &  9,456     \\ \hline
with considered filters  &   877,630    &  8,617     \\ \hline
with at least 25 measurements  &   855,613    &  5,142     \\ \hline
\end{tabular}
\end{table*}

Early efforts in the 20$^{\rm{th}}$ century to understand the mechanism and origin of supernovae focused on classifying these events based on their spectroscopic properties \citep{Minkowski1941,Turatto03}. Type Ia supernovae (SNe Ia) were defined by their strong singly-ionized silicon absorption features and the absence of hydrogen lines. These explosions have since been shown to be consistent with the runaway thermonuclear burning and disruption of carbon-oxygen white dwarfs, based on their observed energetic and nucleosynthetic yields \citep{HN00}.  The ability to calibrate SN Ia luminosities empirically, based on their decline times \citep{Rust74,Pskovskii77, Phillips1993}, or on a timescale stretch \citep{Goldhaber01}, permitted the use of SNe Ia as luminosity distance indicators, rendering them an integral part of the ``cosmic distance ladder'' and revealing the accelerating expansion of the Universe \citep{Riess1998, Perlmutter1999}. The evolutionary channel which causes these white dwarfs to explode, however, remains mainly unknown \citep{MMN14}. Multiple distinct sub-classes of Type Ia supernovae, varying in inferred total luminosity, decline rate, and chemical abundance yields, suggest multiple channels play a role, but connecting these observed classes to theoretical models is a key, ongoing theoretical challenge \citep[e.g.,][]{Ruiter2020}.

The advent of dedicated, wide, high-cadence transient surveys, culminating soon with the Vera Rubin Observatory's Legacy Survey of Space and Time \citep[LSST,][]{LSSTScienceBook}, easily outstrips the community's existing capacity to follow-up event triggers with time-intensive spectroscopy. There is now a pressing need to be able to classify transients based on time-series photometric data (``light curves'') alone. The concurrent exponential growth of machine learning presents a natural approach, and has become an essential tool in the transient community \citep{PLAsTiCC2019}, including the classification of supernovae \citep{DRACULA, PELICAN, Brunel2019, SuperRaenn}. 

In the following, we present a framework for exploring Type Ia supernovae based on their light curves and environmental properties alone, using a dimensionality reduction method known as an implicit rank minimizing autoencoder \citep{IRMA} to guide the interpretation. 

In $\S$\ref{sec:data}, we discuss the Open Supernova Catalog \citep{OSC} and our processing and feature engineering of the available photometric data therein. In $\S$\ref{sec:methods}, we briefly put our work in the context of other efforts to automate the classification of supernovae, before describing the implementation and training of our implicit-rank minimizing autoencoder in detail, and include further discussion of our use of the concept of mutual information in $\S$\ref{sec:mutual}. In $\S$\ref{sec:results}, we identify one of our resulting latent parameters as being strongly correlated with the redshift, and use principal component analysis (PCA) to ``de-redshift'' our other latent variables. We find that three such variables alone describe $\sim$95\% of the variance in our sample, and provide a clear delineation between over-luminous SN 91T events and under-luminous 91bg events. 

Finally, we summarize our results in $\S$\ref{sec:conclusions} and discuss prospects for future improvements and further searches for structure within existing supernova classes.

\section{Data Preparation}
\label{sec:data}

The Open Supernova Catalog is a freely-available, online repository of photometric and spectroscopic supernova data, spanning wavelengths from radio to X-rays, as well as related metadata where available  (e.g., host offset, redshift) \citep{OSC}. At the time of writing, it was comprised of 82,863 individual entries for both supernovae as well as related transients and candidate events. We began our analysis by downloading the entire Open Supernova Catalog via its standard API\footnote{\url{https://github.com/astrocatalogs/OACAPI}. We note while the Open Supernova Catalog effort has been discontinued, the data is still present on the repository.}.

This provided us with 22,528 claimed supernova events. Of these, 14,460 events had photometric data and reasonable values for the event time (MJD), magnitudes, and photometric errors; 9,456 events had complete host information available \citep[drawn from GHOST,][and other sources therein]{GHOST}. The data is heterogeneous, and required selection cuts to ease the analysis, removing events with:

\begin{enumerate}
    \item No photometric data points available in the following 16 bands: 
    $B,V,R,I$, $u,g,r,i,z$, $u',g',r',i',z'$.
    \item SNR $\leq 5$ in all photometric datapoints.
    \item $\leq 25$ measurements for a given event.
\end{enumerate}

After applying the selection cuts, 5,142 events with 855,613 measurements remained. Details of the effect of each cut are shown in table \ref{tab:datacuts}. We further select all events that were exclusively identified as Ia (as some events have  multiple, different claimed types from different sources), often from spectroscopy. This leaves us with 3,640 unambiguous Ia events which have a combined 681,462 measurements. These Ia events are then labelled as \texttt{91T, 91bg, 02cx}
depending on their claimed type if known, and as \texttt{others} if unknown.\footnote{The dataset is publicly available at \url{https://www.canfar.net/citation/landing?doi=23.0034}}

As we will further show, the supernova subtype labels are not used in all steps of the analysis. Following this downselect and re-labelling of the dataset, we then perform the following operations in order to prepare the sample for further analysis:

{\bf Convert magnitude to flux:} The magnitudes $\mathbf{m}$ and uncertainties $\mathbf{\sigma_m}$ are then converted to flux {$\mathbf{f}$ and its uncertainty $\mathbf{\sigma_f}$}. The missing uncertainties are manually set to a constant value, $0.2$ mag \citep{CarrollOstlie}:
\begin{equation*}
    \mathbf{f} = 10^{\frac{22.5-\mathbf{m}}{2.5}}\,\, ,
    \mathbf{\sigma_{f}} = \frac{\ln 10}{2.5} \times \mathbf{f} \times \mathbf{\sigma_{m}}
\end{equation*}

{\bf Data splitting:} Once the data is transformed into flux, we split it into train and test sets. We experiment with two split ratios: a balanced 50\% training to 50\% testing, and a skewed 70\% training to 30\% testing. Initially, we adopted a 50-50 split, aiming to increase the test set sample size, given the limited availability of samples with labeled subtypes. Subsequently, to align more closely with prevalent community standards, we adjusted our approach to the 70-30 split. 

{\bf Data Augmentation:} We augment our dataset artificially with a procedure similar to \texttt{ParSNIP} \citep{ParSNIP}. This is intended to make the trained model more robust against artifacts in the data such as noise, missing values, the uncertainty in determining the time at maximum, etc. The data is augmented by multiple degenerated copies of the same event, while ensuring to preserve the same physical properties. It helps the training process to focus on learning the intrinsic property of the events instead of focusing on the observational or instrumental effects. The following transformations are used to augment the training set, applied randomly in no particular order:
\begin{enumerate}
    \item shift the date of each event, by sampling a time shift $s\sim \cal{N}$(0, 20) days. This transformation intends to simulate an uncertainty in predicting the reference date.
    \item drop a fractional $p$ percent of light curve points, where $p\sim \cal{U}$(0, 0.5), in order to mimic the randomness observations of observation sampling.
    \item scale the amplitude of each event by log-normal $\mathbf{Lognormal}$(0, 0.5). This simulates the multiplicative noise.
    \item add noise to each observation according to $\cal{N}$(0, a), where $a\sim\mathbf{Lognormal}(\log(b), 1)$, $b\sim\mathbf{Lognormal}(-4, 1)$ This simulates the heteroscedacity nature of the data.
\end{enumerate}

\textbf{Pre-processing with Gaussian Process Regression}: light curve data are irregularly sampled in both time and wavelength dimensions and with a wide range of measurement uncertainties. With a procedure very similar to \cite{Avocado}, we model the light curves with Gaussian Processes (GP) regression with a Mat\'ern 3/2 kernel. The smooth modelling enables re-interpolation of the irregular light curves into a grid that is evenly spaced in time and wavelength, together with an estimated uncertainty. We choose to sample the modelled GP light curves onto a grid of 300 sidereal days with 13 bands. The resulting grid can be thought of as a light curve image (LCI), similar to previous supernova studies \citep{PELICAN,SuperRaenn}. The width of the LCI represents a time span from the -149th day relative to the date of maximum brightness up to the 150th day; and the height represents 13 equally spaced wavelengths ranging from 3542.19\AA\ to 10506.54\AA. Each LCI contains two layers, the first layer represents the converted flux value, and the second contains its corresponding the predicted uncertainty from the GP interpolation. Each LCI is then normalized into the range of [0, 1] by dividing through by its maximum value. A lower bound of 0.01 is also set for the uncertainty layer to compensate for the underestimated predicted uncertainty.

\begin{figure}
\begin{center}
\includegraphics[width=3in]{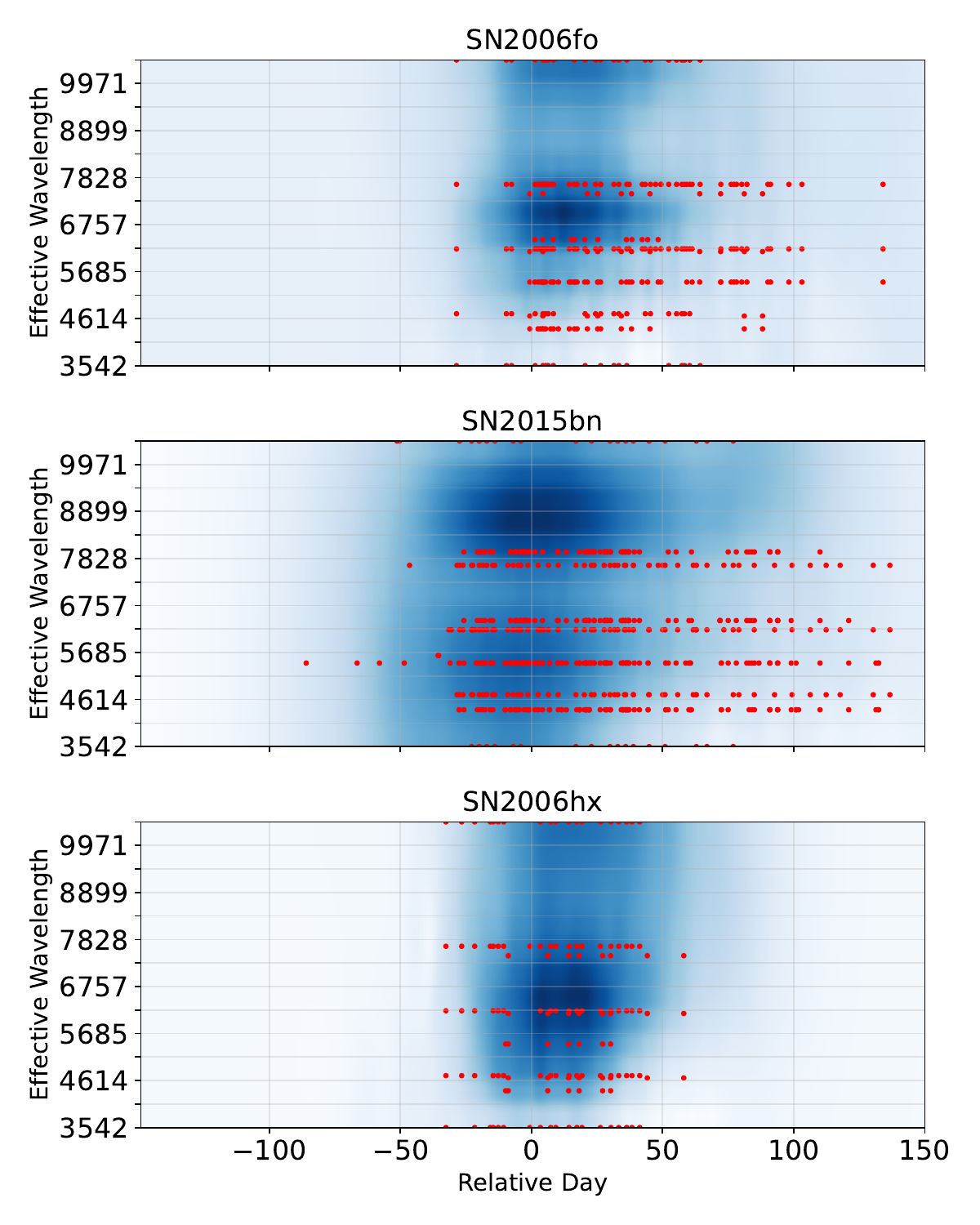}
\end{center}
\caption{\label{fig:sampleLCI}Examples of light curve images (LCI). The red dots represent the location of original data points. The color of the background indicates the normalized flux at a given time and wavelength, which is not necessarily aligned with the density of the original data points.}
\end{figure}

\begin{figure}
\begin{center}
\includegraphics[width=3in]{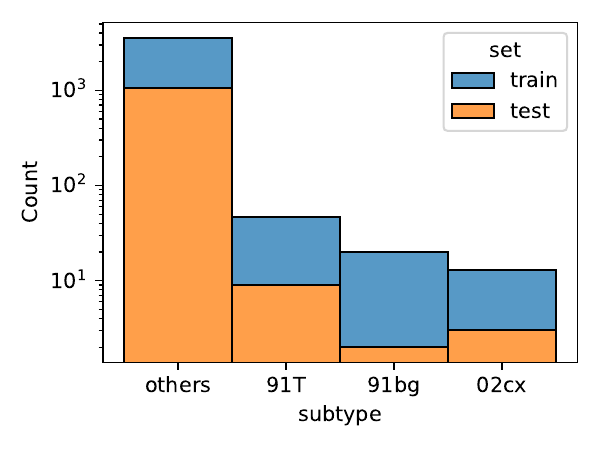}
\end{center}
\caption{Distribution of SNe Ia event classes in the training and test sets. \label{fig:hist_classes}}
\end{figure}

\begin{figure}
\begin{center}
\includegraphics[width=3in]{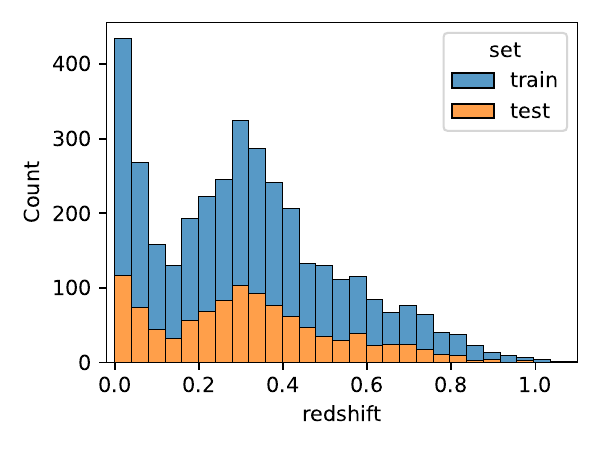}
\end{center}
\caption{Distribution of event redshifts in the training and test set. \label{fig:hist_redshift}}
\end{figure}

\begin{figure}
\begin{center}
\includegraphics[width=3in]{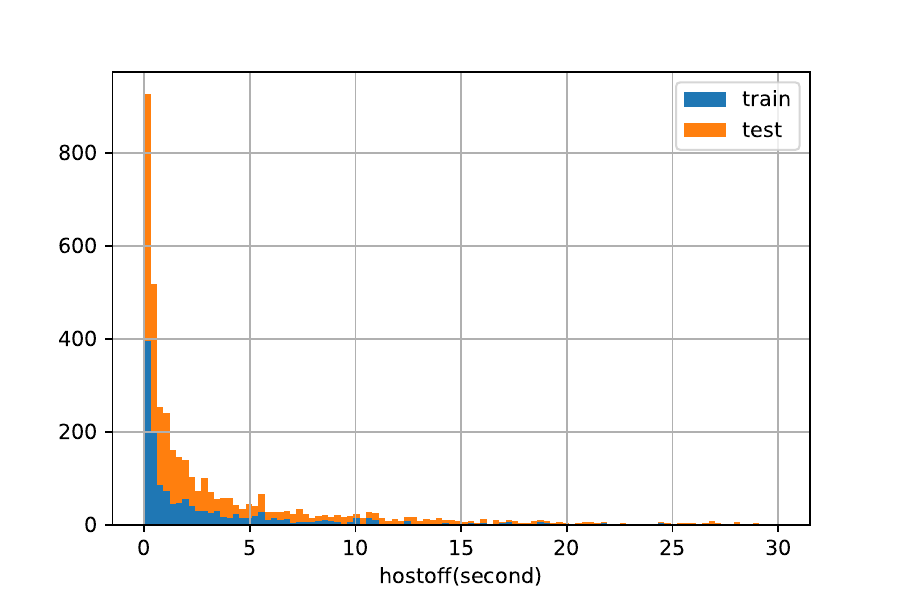}
\end{center}
\caption{histogram for host galaxy offset. \label{fig:hist_hostoffset}}
\end{figure}

\section{Machine-learning classification of type Ia supernovae}
\label{sec:methods}

\subsection{Related Studies}

Previous studies have investigated the application of machine learning and deep learning architecture to photometric classification. In \texttt{PELICAN}\citep{PELICAN}, the team presents a contrastive deep learning architecture with a supervised module that infers class label directly from a LCI as well. The LCI is a single layer image without the uncertainty. The team performed photometric classification using an architecture that is similar to the ones used in image classification. To deal with the missing values presented in the LCI, multiple modules and regularizers are part of the architecture such that the network is capable of inferring class label from the incomplete images. The architecture was tested on simulated light curves of the LSST Deep Fields, and was able to detect 87.4\% of Type Ia supernovae with a precision higher than 98\%. 

\texttt{SuperRAENN} \citep{SuperRaenn} presents a semi-supervised learning method that involves three major steps in the pipeline that are similar to the ones presented here: 1) Gaussian Process (GP) interpolation on the multi-band lightcurves; 2) a recurrent autoencoder to encode a common representation 3) classification on the encoded vectors with a supervised random forest \citep{breiman_random_2001}. The completed pipeline has an accuracy of 87\% across five SN classes (Type Ia, Ibc, II, IIn, SLSN-I) on PS1-MDS. 

\texttt{Avocado} \citep{Avocado} also shows a similar traditional approach, consisting of: 1) Gaussian Process (GP) interpolation on the multi-band light curves; 2) data augmentation of the light curves and different redshift; 3) extraction of physical features from the light curve GP fit 4) classification on extracted features with a gradient-boosted decision tree.  This algorithm was the best-performing among the 1094 models considered in the blinded phase of the Photometric LSST Astronomical Time-Series Classification Challenge \citep{PLAsTiCC2019}, scoring 0.468 on the organizers’ logarithmic-loss metric with flat weights for all object classes in the training set, and achieving an AUC of 0.957 for the classification of SNe Ia.

A particularly powerful approach is to combine photometric light curves with the enormous complementary information in supernova spectra, where available. Adopting a machine learning approach to spectral classification of supernovae and their sub-types was pioneered by \cite{Modolo2015}, using artificial neural networks, and \cite{Sasdelli2015}, using principal component analysis. \cite{Sasdelli2017} demonstrated that even with a very small sample of events it was possible to delineate different explosion models within their principal component space. Subsequent efforts have expanded a hybrid approach combining LCI with spectra and other metadata such as host galaxy characteristics \citep[e.g.,][]{DRACULA, Brunel2019, Muthukrishna2019, Santos2020, Chen2020, ArantesFilho2021}. Taken together, these studies have shown the great promise for a hybrid approach to refine the characterization of Type Ia supernova sub-types.

\subsection{Using autoencoders}

Our dataset is composed of multi-band supernova light curves, and where available their contextual data including the host galaxy environment, redshift, and Milky Way line-of-sight extinction. Our goal is to be able to explore the diversity of supernovae without initially imposing a pre-defined classification scheme, to eventually associate physical models to identified similar data. For this exploration, we need to homogenize and reduce the dimensionality of the data set. Reduction to two dimensions for visualization enables better interpretation. Dimensionality reduction is not required for a classification objective but having a single consistent and flexible encoding of the data renders the analysis and exploration more tractable. The light curves data in this case are also quite heterogeneous, with multiple bands and time gaps between observations. Here, we learn an encoded version of all the data of one supernova event, with the encoding framed within the neural network autoencoder.

In their traditional form \citep{bank2023autoencoders}, autoencoders provide a nonlinear mapping of input data into an encoded representation, capturing enough information to be able to reconstruct the input data with minimal loss in the least-square sense. This initial form of self-supervision has been shown to be flexible and has been utilized for a wide variety of applications. Several variants of autoencoders exist, and have previously been explored for supernova classification and analysis (see $\S$3.1 above). For our analysis, we explored several variants of autoencoders suited for a minimally constrained encoding, capable of capturing the diversity of the current supernova dataset. A popular variant is the Variational Autoencoder (VAE) \citep{VariationalAutoencoder}, that enables resampling from a Gaussian distribution on the latent space, thereby becoming more compact for generative modelling. Here we explicitly keep the latent space more flexible, and focus mode on the encoder part. We regularize the encoder with an implicit rank minimizing autoencoder which we detail in the following section.

\subsection{Implicit Rank Minimizing Weighted Autoencoders}\label{IRMAE}

Autoencoders are represented by neural networks, a composition of linear functions each activated by non-linear activations. By keeping the last composed functions (layers) before encoding the data, \cite{jing2020implicit} have shown an autoencoder implicitly minimizing the rank of the covariance matrix for latent variables. The latent variables of such a network summarize the variation of the input data with a minimal number of significant uncorrelated axes.

Here we exploit this simple version of the autoencoder motivated by the fact that rank-minimizing decomposition typically extracts more informative and less redundant features. Let $\mathbf{x}_i$ represents all pixels of a LCI; the optimal weights of the network $f_\theta$ are then found by minimizing the cost function:
\begin{equation}
    \hat{\theta} = \argmin_\theta \frac{1}{N}\sum_{i=1}^{N}\left(\frac{\mathbf{x}_i-f_{\theta}(\mathbf{x}_i,\mathbf{m}_i)}{\sigma_{\mathbf{x}_i}}\right)^2
\end{equation}
where $\theta$ denotes the weights of the autoencoder network; $N$ denotes the total number of LCI samples; $\sigma_{\mathbf{x}}$ denotes the uncertainty of $\mathbf{x}$, and $\mathbf{m}$ denotes the associated metadata. The full network is then composed of an encoder and decoder $f=e\circ d$, where the encoder can be thought of a succession of non-linear and rank-minimizing linear layers $e=e_{\mathrm{NL}}\circ e_\mathrm{L}$. The losses are weighted according to their uncertainties similar to a Gaussian log-likelihood, to encourage the network to reconstruct a light curve that well approximates the observations with low uncertainty while giving the network flexibility on observations that have high uncertainty. We note that when the metadata encoding is enabled, the architecture is not strictly an autoencoder as there is no decoder to reconstruct the metadata.

When the LCI training set is fed into the network, it is converted into a low-dimensional vector representation by the encoder. The vectors then pass through a series of linear layers (rank-minimizing block), further reducing the rank of their covariance matrix. During training, the decoder uses a vector $\mathbf{z}$, concatenated with available metadata (see above), to reconstruct the input LCI. After training, the encoder, along with the rank-minimizing block, can be used to perform the dimensional reduction.

\begin{figure}
\begin{center}
\includegraphics[width=2in]{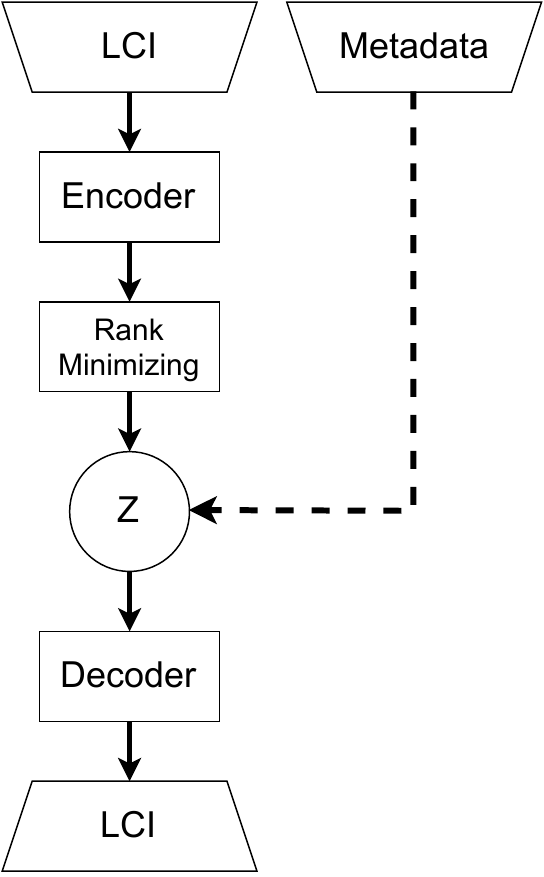}
\end{center}
\caption{\label{fig:IRMAE}Overview of IRMAE structure. $\mathbf{z}$ is the vector representation of the LCI, denoted latent variables. The metadata can be composed of environmental factors e.g. redshift, colors of the host galaxy and angular separation.}
\end{figure}

\section{Mutual Information Analysis}
\label{sec:mutual}
Before proceeding to our analysis, we introduce one more concept which will allow us to compare our derived latent variables with observed quantities. Mutual Information is a non-parametric quantity to measure the dependency of two random variables. Let $(X, Y)$ be a pair of random variables, $P_X(x)$ and $P_Y(y)$ be the marginal probability of $X$ and $Y$, and $P_{XY}(x, y)$ be their joint probability. The mutual information\citep{cover1999elements} of them is given as:
\begin{equation}
    I(X;Y) = D_{KL}(P_{XY}||P_XP_Y)
\end{equation}
where $D_{KL}$  is the Kullback–Leibler divergence\citep{joyce2011kullback}, which measures the similarity of two distribution, in this case, the joint distribution and the product of two marginal probabilities. The value of mutual information is non-negative. The mutual information is zero when $X$ and $Y$ are independent and the value increase along with dependency. The procedure to calculate mutual information depends on whether the variables are continuous or discrete. \citep{ross2014mutual}

\section{Results}
\label{sec:results}

\subsection{Parameter dependence on redshift}
\begin{figure}
\begin{center}
\includegraphics[width=3.5in]{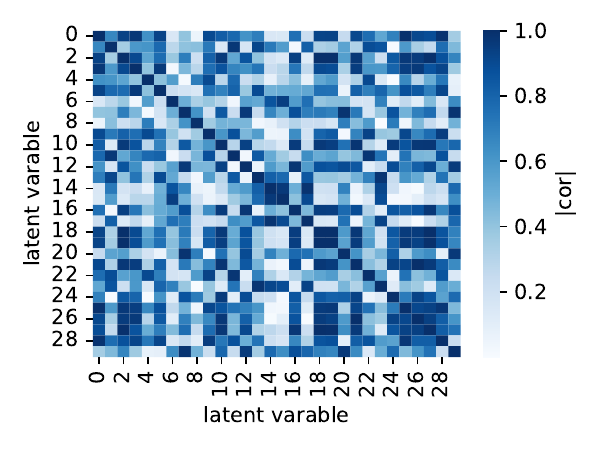}
\end{center}
\caption{\label{fig:latent_var_absCorr}Correlation matrix of latent variables of the autoencoder. Most latent variables are highly correlated. This means that there are redundant information among the latent variables.}
\end{figure}

\begin{figure}
\begin{center}
\includegraphics[width=3in]{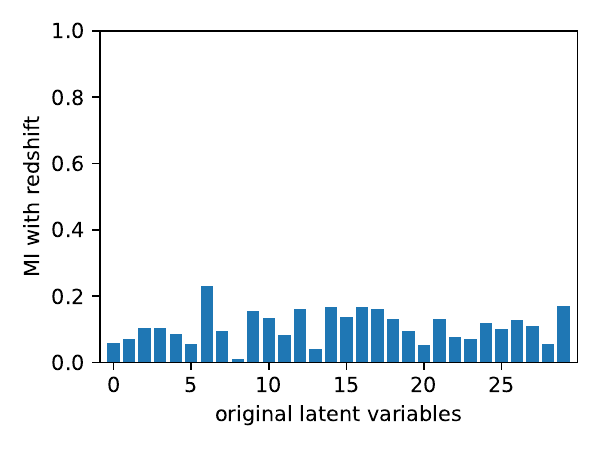}
\end{center}
\caption{\label{fig:latent_redshift_MI}Mutual information between latent variables and redshift. Most of the latent variables also has moderate dependency with redshift.}
\end{figure}

After reducing the dimensionality of the LCI, we obtain 30 latent variables, of which the covariance has a low rank, meaning there are only a small number of linearly independent components in the space of latent variables contributing significantly to the overall variability. However, the latent variables are not necessarily disentangled. The significant linearly independent components can be a linear combination of the latent variables. This is reflected in the correlation matrix in figure \ref{fig:latent_var_absCorr}. Moreover, the latent variables are also entangled with redshift, with most showing a moderate degree of correlation with redshift, seen in Fig. \ref{fig:latent_redshift_MI}.

\subsection{De-redshifting using PCA}
The two effects mentioned above make analyzing the variability of latent variables difficult. To de-correlate the latent variables, while removing/mitigating the effect of redshift, we concatenate the latent variable with the redshift value and apply a principal component analysis (PCA) on the concatenated variable. In this way, we transfer the concatenated variable to a new space $S$, whose axes are not correlated.

\begin{figure}
\begin{center}
\includegraphics[width=3in]{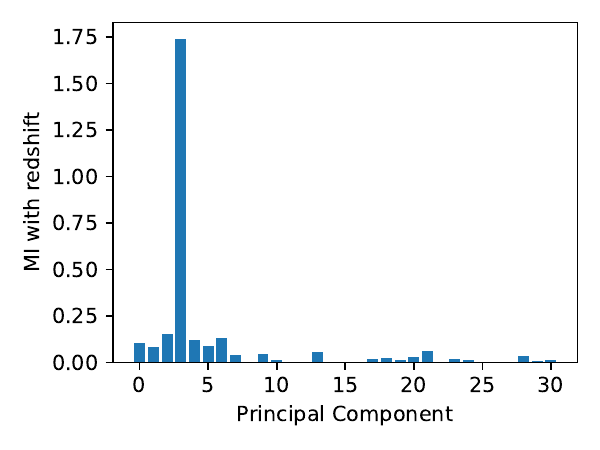}
\end{center}
\caption{\label{fig:PC_redshift_MI}Mutual information between principal components and redshift. Two principal components have noticeable dependence with redshift, while the other components have less dependency. This suggests the information about redshift is concentrated mostly in this two axes.}
\end{figure}

We again check the correlation between PCA transformed latent variable and the value of redshift. We find that only a small number of transformed latent variables are correlated to redshift, and only one of them shows a strong correlation (see figure \ref{fig:PC_redshift_MI}). We identify this axis as the one corresponding to the redshift effect and remove it from the space $S$.

\subsection{Independent Component Analysis}
By applying the PCA, latent variables are transformed to a new space where all axes are not correlated to each other. However, no correlation implies independence only when the variables are Gaussian, which is not always the case. Independent component analysis (ICA) is a method that finds independent axes by pushing the assumption of non-normality. It assumes the observed signal is a mixture of independent non-Gaussian source signal. Suppose we have an observation $x$ that is a mixture of $n$ source signal $s_i$ weighted by $a_i$.

\begin{equation}
    x = \sum_{i=1}^{n} a_i s_i
\end{equation}

\begin{figure}
\begin{center}
\includegraphics[width=3in]{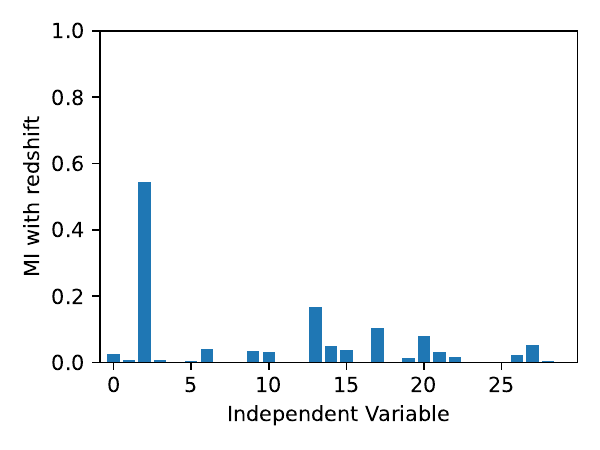}
\end{center}
\caption{\label{fig:Ivar_redshift_MI}Mutual information between principal components and redshift. The overall dependency with redshift is much lower in the new independent variables comparing to the old ones in Figure \ref{fig:latent_redshift_MI}.}
\end{figure}

The central limit theorem establishes that, in many situations, the summation of independent random variables are asymptotically normal. This means that the observation $x$ should be more similar to the Gaussian distribution than the individual source signal $s_i$. ICA uses this principle to find the source signal by maximizing the non-normality of the source signal under the constraint that all source signals are uncorrelated. In our case, we consider the entangled latent variable as the observed variable $x$ and the disentangled variables as the source signal $s$. We find applying ICA on top of the previous procedure further mitigates the effect of redshift on the variables, as shown in Fig. \ref{fig:Ivar_redshift_MI}.

\subsection{Separating SN Ia subclasses}
\begin{figure}
\begin{center}
\includegraphics[width=3in]{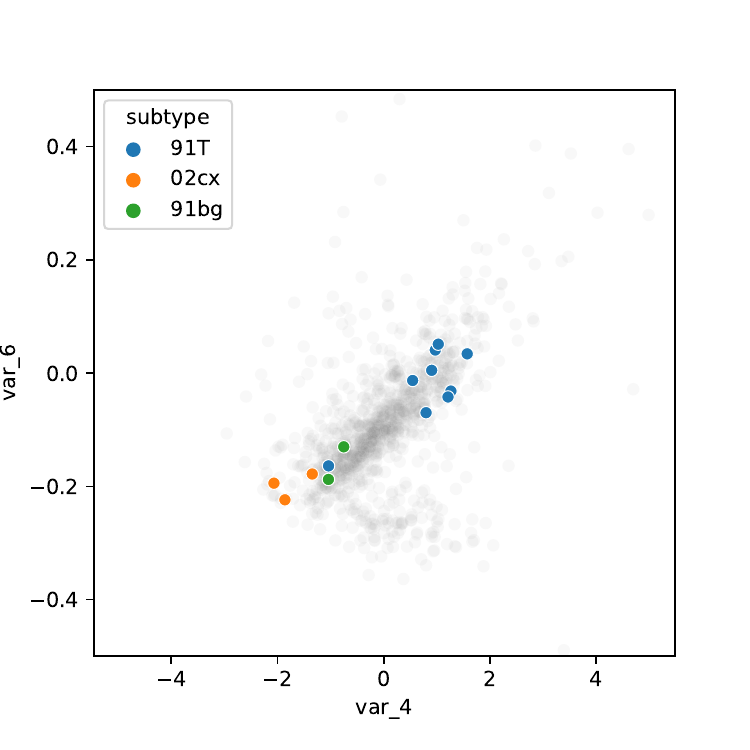}
\end{center}
\caption{\label{fig:jointplt_selected_vars}PC1 vs PC5. The transparent gray points are unlabeled data. In these hand-picked principal axes, we can observe the separation between 91T-like and the other two.}
\end{figure}

\begin{figure}
\begin{center}
\includegraphics[width=3in]{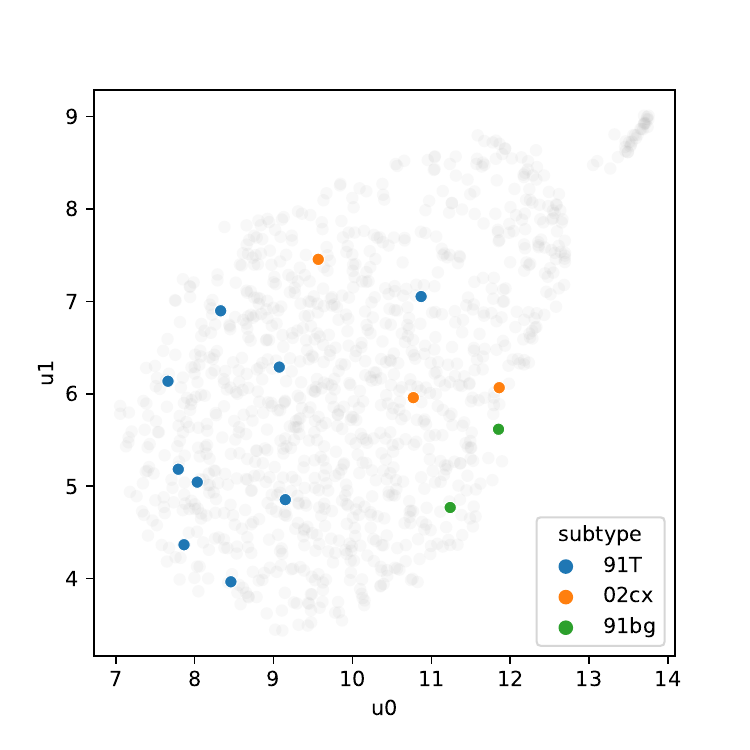}
\end{center}
\caption{\label{fig:jointplt_UMAP}UMAP for ICA components. The transparent gray points are unlabeled data. This shows the visualization of all independent components. We got the same observation as Figure \ref{fig:jointplt_selected_vars}.}
\end{figure}

\begin{figure}
\begin{center}
\includegraphics[width=3.3in]{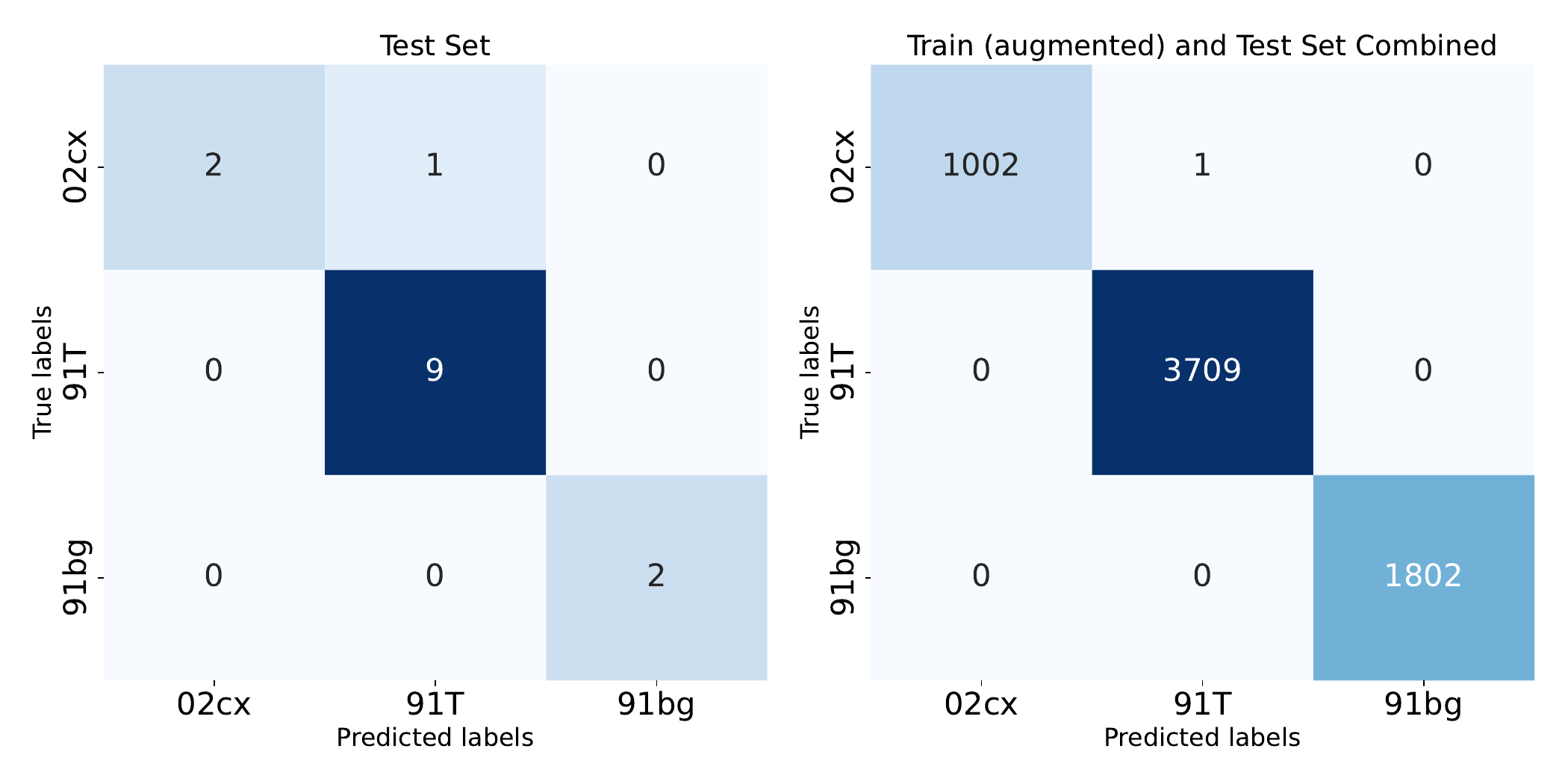}
\end{center}
\caption{\label{fig:confusionMatrix}Confusion matrix for the classification on test and combined sets. A random forest classifier is trained on the encoded train set and applied on both test and the combined sets.}
\end{figure}

We manually inspect the rest of the transformed latent variables. We find that two variables are particularly useful for separating some Type Ia supernova sub-classes, in particular delineating luminous 91T-like events and  underluminous 91bg-like and 02cx-like events. In Fig. \ref{fig:jointplt_selected_vars}, we observe that 91T-like SNe are mostly located in the upper right part of the plot, well separated from 02cx-like. We also find similar behaviour in the ICA map (see Fig. \ref{fig:jointplt_UMAP}), where 91T-like are accumulated in lower left of the map while 91bg-like and 02cx-like are located in the upper right. Further, we train a Random Forest classifier to verify these observations (Fig.\ref{fig:confusionMatrix} and Table \ref{tab:classification}). Interestingly, we do not find that either the over-luminous 91T or the under-luminous 91bg populations are separated from the main body of ``other'' supernovae, including all ``normal'' SNe Ia in the sample.

Note however that we do find that there are three 91T-like supernovae located in the 91bg-like and 02cx-like region in our parameter space. Upon further scrutiny, we found that these outliers are in fact the same supernovae in both maps, namely SN2005eq, SN2016isl, and SNF20080522-000. To assess what this means for our analysis, we next turn to the physical interpretation of the mean light curves within each class and contrasting these with the outlier events.

\begin{table}[]
\caption{Classification on parameter space\label{tab:classification}}
\begin{tabular}{r|r|r|}
\cline{2-3}
\textbf{}                       & \textbf{Test Set} & \textbf{Combined Set} \\ \hline
\multicolumn{1}{|r|}{Accuracy}  & 0.928571          & 0.999846              \\ \hline
\multicolumn{1}{|r|}{Precision} & 0.935714          & 0.999847              \\ \hline
\multicolumn{1}{|r|}{Recall}    & 0.928571          & 0.999846              \\ \hline
\multicolumn{1}{|r|}{F1 Score}  & 0.923308          & 0.999846              \\ \hline
\end{tabular}
\end{table}

\subsection{Mean light curves and physical interpretation}

One possible explanation for the outlier events in our sample may be missing observations in I band, resulting in over-smoothing of the light curves. To investigate this, we turn to checking the raw and imputed light curves for these three events. 

\begin{figure*}
\begin{center}
\includegraphics[width=0.8\textwidth]{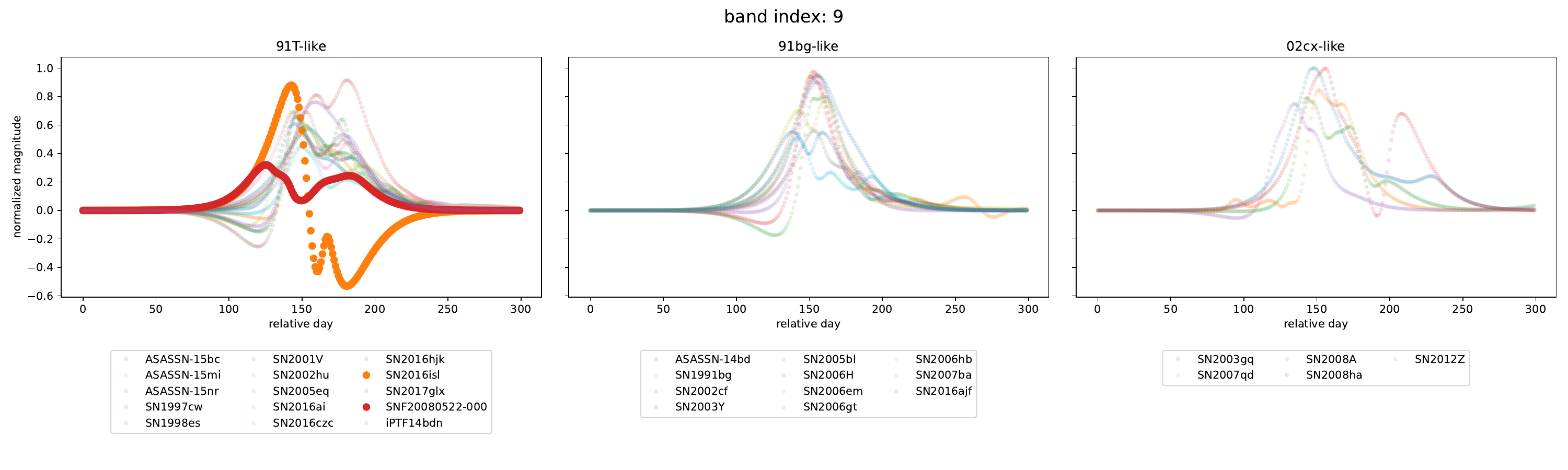}
\end{center}
\caption{\label{fig:GP_b8765}This, along with, Figure \ref{fig:GP_b4702_hitlight05eq} and \ref{fig:GP_b9926_hitlight05eq} shows the GP regression curves for each subtype in selected bands of LCI. These plots intended to show the pattern of LCI for each subtype and try to explain the three "misplaced" 91T-like events. The events with solid line are those in question. As shown in this plot, two of the events has artifacts in the higher wavelength band (8765 Angstroms) of the LCI.}
\end{figure*}

Among them, SN2016isl and SNF20080522-000 are missing observations in the $I$ band. However, instead of an over smoothed light curve, we found an interpolation artifact, such that the light curve in the higher wavelength band has distorted shape in longer wavelength, see Fig. \ref{fig:GP_b8765}.

Unlike the other two misplaced events, SN2005eq has observations in the I band, and the imputed curves also exhibit the double-peaks. This implies that the autoencoder may not discriminate the light curves based on the signature double peaks, but other factors.

\begin{figure*}
\begin{center}
\includegraphics[width=0.8\textwidth]{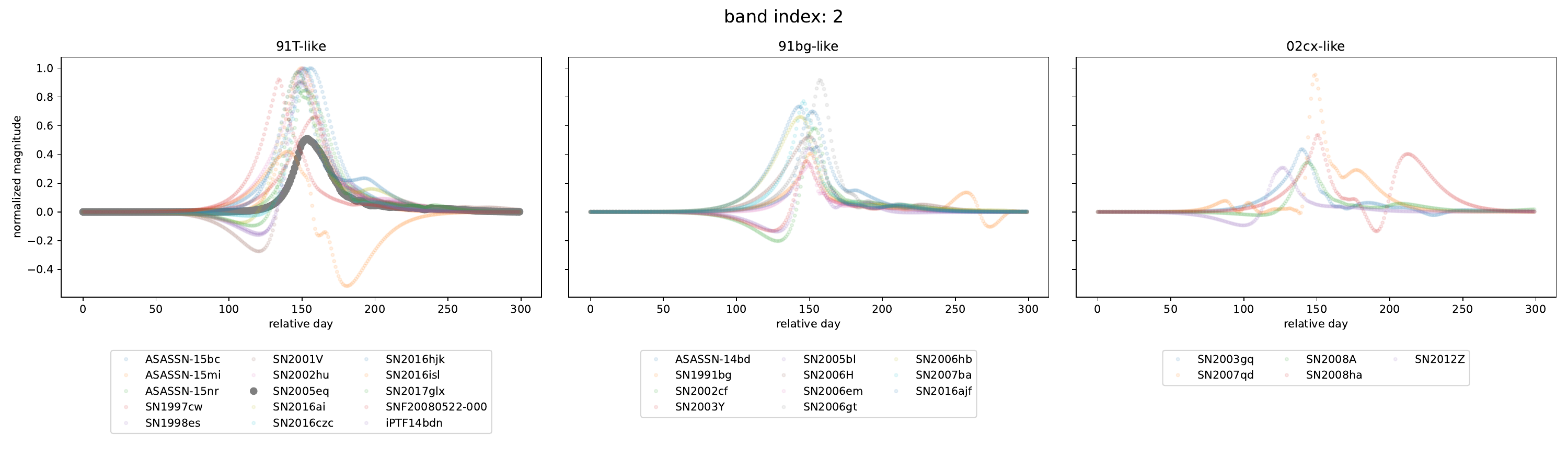}
\end{center}
\caption{\label{fig:GP_b4702_hitlight05eq}This, along with Figure \ref{fig:GP_b9926_hitlight05eq} shows the problem of SN2005eq. The plots here show the bands where SN2005eq is extrapolated, which is similar to 91bg-like in relative magnitude.}
\end{figure*}

\begin{figure*}
\begin{center}
\includegraphics[width=0.8\textwidth]{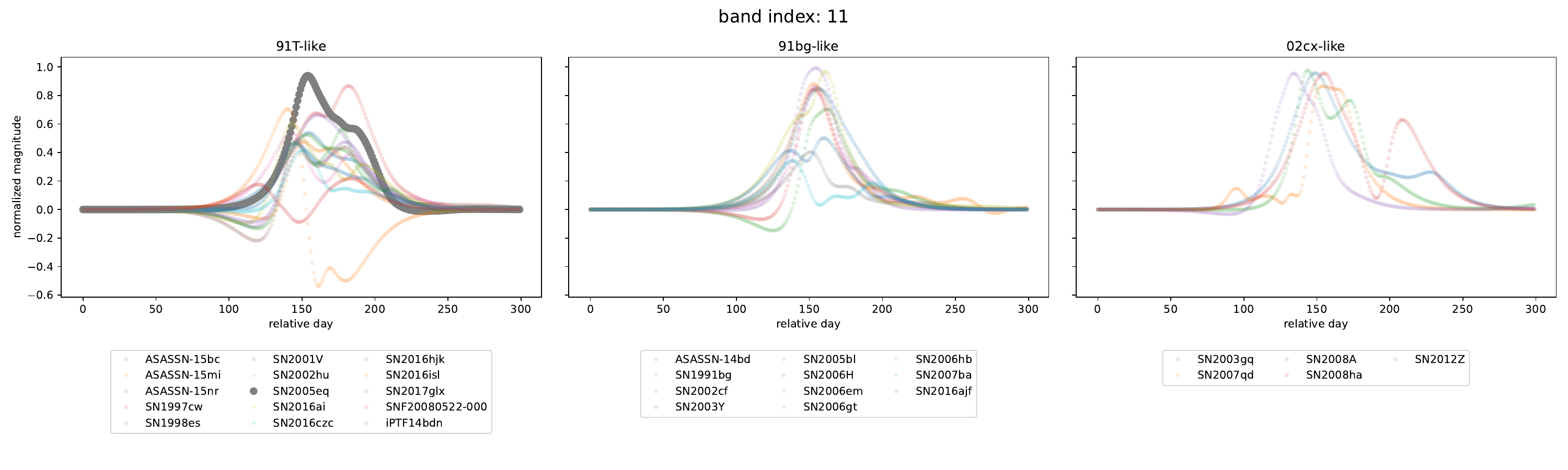}
\end{center}
\caption{\label{fig:GP_b9926_hitlight05eq}GP regression curves in band 11.}
\end{figure*}

In our samples, the imputed light curves for 91T-like supernovae typically have higher relative brightness in shorter wavelength band, comparing to 91bg-like supernovae (Figs. \ref{fig:GP_b4702_hitlight05eq} and \ref{fig:GP_b9926_hitlight05eq}). Although the imputed light curve of SN2005eq peaks at the band with the longest wavelength, this is only due to the extrapolation by the GP; we found that the original light curve does not. 

Contrary to physical modelling, data-driven modelling techniques struggle with extrapolation and GP is no exception. GP regression can be considered as the average of random walks with specified covariance function conditioned in the context points. Without the constraint of the context points, the prediction results of GP regression are usually incorrect. As a result, due to the missing observations in longer wavelength bands, the imputed light curves for SN2005eq are peaked at much longer wavelengths than the actual light curves. This may explain why SN2005eq is misplaced in the 91bg-like region.

\section{Discussion}
\label{sec:Discussion}
The dataset includes information about the supernova host galaxies, such as their color, redshift, and angular separation from the supernova. However, in our final model, we use only the host galaxy redshift as metadata, since we found that including the color and angular separation of the host galaxies did not significantly improve our results. The current method suffers from extrapolation problems during the Gaussian process imputation. Due to the heterogeneous nature of the dataset, it is challenging to avoid cutting large amounts of data while maintaining quality. The present goal is to develop a method to address this challenge such that we can use one model to describe all data from difference sources (with different bands, time range, and cadence). Ideally, by weighting the loss function of the network with uncertainty, the autoencoder can focus on the parts of LCI where data exist and summarize the physical information of the data into an abstract representation.

However, the extrapolation problem in LCI reveals that the predictive uncertainty from Gaussian process regression may not properly reflect the error. The predictive uncertainty can still be small in extrapolated areas of the LCI. This causes the autoencoder to include the imputation artifacts in its learning and inferring process, polluting the abstract representation.

In order to mitigate the imputation issue, we propose two solutions. The first is to apply stricter conditions in the quality cut, such that the data become easier to impute; the second one is to adjust the setting of GP regression that is tailored to our dataset but it will not guarantee that generalizes to other datasets. Unfortunately, either approach would hinder our ability to construct a universal model that describe compose data with multiple sources, however this is an area for future work.

\section{Conclusions}
\label{sec:conclusions}

Our approach involved an implicit rank-minimizing autoencoder, which effectively compressed the LCI parameter space from the Open Supernova Catalog, achieving a reduction to 30 latent variables. While this number might not seem remarkable, the true strength of the method lies in its ability to condense the rank of the latent space, rather than just its dimensionality. Using PCA, we were able to further identify one principal component as strongly correlated with the redshift (confirmed using ICA). This allowed us to de-redshift our parameter space, and identify three latent variables which otherwise accounted for $\sim$95\% of the variance of our sample. Two such components provided an interesting delineation of Type Ia supernova subclasses, with confirmed 91T events lying to one side of the resulting mapping and 91bg and 02cx events lying at the other extreme, and with only three outlier events for which we were able to identify to causes. Notably, while the overluminous 91T events and the underluminous 91bg and 02cx events are well-separated in this space, neither is particularly distinguishable from the overall population of ``other'' Type Ia supernova events in the sample, including all `normal' events. In the future, directly combining this photometric data with spectroscopic observations, where available, may provide a powerful means to investigate the physics underlying different Type Ia supernovae subclasses and their possibly heteraogenous origins. 
\bibliography{sncluster}{}
\bibliographystyle{aasjournal}

\end{document}